\documentstyle[aps]{revtex}
\begin{document}
\draft
\title{Phase Transition in Conformally Induced Gravity with Torsion}
\author{Jewan Kim$^{*}$, C. J. Park$^{*}$ and Yongsung Yoon$^{\dagger}$}
\address{$^{*}$Department of Physics, Seoul National University \\
Seoul 151-742, Korea \\
$^{\dagger}$Department of Physics, Hanyang University \\
Seoul 133-791, Korea \\
E-mail: atamp@hepth.hanyang.ac.kr}
\maketitle
\begin{abstract}
We have considered the quantum behavior of a conformally
induced gravity in the minimal Riemann-Cartan space.
The regularized one-loop effective potential considering
the quantum fluctuations of the dilaton and the torsion fields in
the Coleman-Weinberg sector gives a sensible phase transition for
an inflationary phase in De Sitter space.
For this effective potential, we have analyzed the semi-classical
equation of motion of the dilaton field in the slow-rolling regime.
\end{abstract}
\pacs{04.62.+v, 11.30.Qc, 98.80.Cq}

\section{Introduction}

Among the four fundamental interactions in nature,  the  two
feeble  interactions   are  characterized  by  dimensional
coupling  constants;  those  are  the Fermi's  coupling  constant
$G_{F}= (300 Gev)^{-2}$ and Newton's coupling constant $G_{N}= (10^{19}
Gev)^{-2}$.

The  interactions  with  dimensional  coupling  constants of
inverse mass dimensions are strongly diverse and nonrenomalizable.
However, from the success of Weinberg-Salam model,
the weak interaction at a fundamental level is actually characterized
by  a  dimensionless coupling constant, and the  dimensional
nature of $G_{F}$ results from a spontaneous symmetry breaking.
Indeed $G_{F}\cong  \frac{1}{v_{\omega}^{2}}$,
where $ v_{\omega}\cong 300$ Gev
is the vacuum expectation  value  of Higgs field.
The weakness of the weak interaction comes from the largeness of
the vacuum expectation value of Higgs field \cite{Weinberg}.

In the light of the above remarks,  it is considerable that gravity  is
also  characterized by a dimensionless coupling constant and
that the weakness of gravity is associated with symmetry breaking
at the high energy scale. Similarly to $G_{F}$,
$G_{N}$ is given by the inverse square of the vacuum expectation  value  of
a scalar  field, dilaton.
It was independently proposed by Zee \cite{Zee} and  Smolin \cite{Smolin}
that the Einstein-Hilbert action,
\begin{equation}
   S= -\int d^{4}x \sqrt{g}\frac{1}{16 \pi G_{N}} R,
\label{hil}
\end{equation}
 can be replaced by the modified action
\begin{equation}
     S =  \int d^{4} \sqrt{g}(-\frac{1}{2}\xi \phi^{2} R +
	  \frac{1}{2} \partial_{\mu}\phi \partial^{\mu}\phi - V(\phi)),
\label{hil1}
\end{equation}
where the coupling constant  $\xi$ is dimensionless.
The potential $V(\phi )$ is
assumed to attain its minimum value when  $\phi = \sigma$, then
\begin{equation}
   G_{N}=\frac{1}{8\pi\xi\sigma^{2}}.
\label{grv}
\end{equation}
 Through  the  spontaneous symmetry breaking, the
symmetric  phase of the scalar field transits to  an asymmetric  phase
of the scalar field.
     On  the  analogy  of the $SU(2) \times U(1)$  symmetry  of  the  weak
interactions, we can consider a symmetry which is broken through
spontaneous symmetry breaking in the gravitational  interactions.
The  most  attractive
symmetry  is  the  conformal symmetry which rejects  the  Einstein-
Hilbert action Eq.(\ref{hil}), but  admits the modified action Eq.(\ref{hil1})
with the specific coupling $\xi = -\frac{1}{6}$ and quartic potential.
We can write   down   a   conformally  invariant  induced gravity action
without   introducing  the  torsion  field.  However, the spontaneous symmetry
breaking
mechanism does not work for the scalar field theory with $\xi = -\frac{1}{6}$
in De Sitter space.

In  the  minimal Riemann-Cartan space,  the vector  torsion  behaves
effectively like a conformal gauge field \cite{Nieh}. The introduction of
this torsion field makes the dimensionless coupling constant in Eq.(\ref{hil1})
free in the conformally invariant induced gravity action.
Therefore, it is necessary to introduce the torsion field into
the conformally induced gravity.
Since we expect that
the  conformal  symmetry  is broken at  very high  energy scale,  it  is
natural to consider the conformal symmetry  with
an inflation scenario \cite{la}. We have investigated the quantum behavior  of
the dilaton and the vector torsion field in the conformally induced gravity.
The one-loop effective potential of this action exhibits  kind of
a phase transition which may be responsible for an inflation scenario
\cite{Tye,Guth,Linde,Albrecht}.

\section{Conformally Induced Gravity in Minimal Riemann-Cartan Space}

In this section we construct a conformally invariant induced gravity action
with torsion field.  Let us start from the condition of
 conformal  invariance  of  the   tetrad postulation
\begin{equation}
       D_{\alpha} e^{i}_{\beta}\equiv\partial_{\alpha}e^{i}_{\beta} +
       \omega^{i}_{j\alpha} e^{j}_{\beta}-\Gamma^{\gamma}_{\beta \alpha}e^{i}_{
       \gamma} = 0,
\label{cor}
\end{equation}
to find how the connection and the torsion behave  under the conformal
transformation,
\begin{equation}
     (e^{i}_{\alpha})'= exp(\Lambda(x)) e^{i}_{\alpha},
     ~~~(\omega^{i}_{j\alpha})' = \omega^{i}_{j\alpha}.
\label{cor1}
\end{equation}
We have used Latin indices for the tangent space and  Greek indices  for the
curved space.
{}From the  above requirement, the
asymmetric affine connection and the torsion which is the antisymmetric part
of the connection transform as follows;
\begin{equation}
          (\Gamma^{\gamma}_{\beta \alpha})'
    =\Gamma^{\gamma}_{\beta \alpha} + \delta^{\gamma}_{\beta} \partial_{\alpha}
    \Lambda,
\label{aff}
\end{equation}
\begin{equation}
          (T^{\gamma}_{\beta \alpha})'=
          T^{\gamma}_{\beta \alpha}+
	  \delta^{\gamma}_{\beta}\partial_{\alpha}\Lambda
        - \delta^{\gamma}_{\alpha}\partial_{\beta}\Lambda,
\label{tor1}
\end{equation}
\begin{equation}
          (T^{\gamma}_{\gamma \alpha})'=
          T^{\gamma}_{\gamma \alpha}+
	  3\partial_{\alpha}\Lambda.
\label{contor}
\end{equation}
Therefore, the  contracted torsion $T^{\gamma}_{\gamma \alpha}$ is effectively
playing the
role of a conformal gauge field.  We can separate the torsion into  two
components;
\begin{equation}
          T^{\alpha}_{\beta \gamma}=
          A^{\alpha}_{\beta \gamma}-
	  \delta^{\alpha}_{\gamma}S_{\beta}+
	  \delta^{\alpha}_{\beta}S_{\gamma},
\label{tor2}
\end{equation}
\begin{equation}
          (S_{\alpha})' =
	  S_{\alpha}+
	  \partial_{\alpha}\Lambda, ~~~
          (A^{\alpha}_{\beta \gamma})'=
          A^{\alpha}_{\beta \gamma} .
\label{vec}
\end{equation}

To  avoid  the unnecessary complexity,  we adopt the  conformally
invariant torsionless condition
\begin{equation}
          A^{\alpha}_{\beta \gamma} \equiv 0.
\label{zero}
\end{equation}
Because this  condition is the conformally invariant  extension  of  the
torsionless condition in Riemann space $ T^{\alpha}_{\beta \gamma}\equiv 0$,
we call this space as the minimal Riemann-Cartan space.
For this space, the affine connection is solved in terms of
$g_{\mu\nu}$ and $S_{\alpha}$;
\begin{equation}
          \Gamma^{\alpha}_{\beta \gamma}=
	  \{^{\alpha}_{\beta \gamma}\}+
          S^{\alpha} g_{\beta\gamma}-
          S_{\beta}\delta^{\alpha}_{\gamma},
\label{cri1}
\end{equation}
Let us define the conformally invariant connection,
$\Omega^{\alpha}_{~\beta \gamma}$;
\begin{equation}
          \Gamma^{\alpha}_{~\beta \gamma}=
          \Omega^{\alpha}_{~\beta \gamma}+
	  \delta^{\alpha}_{\beta}S_{\gamma},
\label{con}
\end{equation}
\begin{equation}
          \Omega^{\alpha}_{~\beta \gamma}=
	  \{^{\alpha}_{\beta \gamma}\}+
          S^{\alpha}g_{\beta\gamma}-
          S_{\beta}\delta^{\alpha}_{ \gamma}-
          S_{\gamma}\delta^{\alpha}_{ \beta}.
\label{con1}
\end{equation}
The curvature tensor of the affine connection $\Gamma^{\alpha}_{\beta \mu}$,
\begin{equation}
R^{\alpha}_{~\beta \mu \nu}(\Gamma)=
\partial_{\mu}\Gamma^{\alpha}_{\beta \nu}-
\partial_{\nu}\Gamma^{\alpha}_{\beta \mu}+
\Gamma^{\alpha}_{\sigma \mu} \Gamma^{\sigma}_{\beta \nu}-
\Gamma^{\alpha}_{\sigma \nu} \Gamma^{\sigma}_{\beta \mu},
\label{cur}
\end{equation}
can be expressed in terms of
$\Omega^{\alpha}_{~\beta\gamma}$ and $S_{\alpha}$  using Eq.(\ref{con});
\begin{equation}
R^{\alpha}_{~\beta \mu \nu}(\Gamma)=
R^{\alpha}_{~\beta \mu \nu}(\Omega)+
\delta^{\alpha}_{\beta}H_{\mu \nu},
\label{cur1}
\end{equation}
\begin{equation}
R_{\alpha \nu}(\Gamma)=R_{\alpha \nu}(\Omega)+H_{\alpha \nu},
\end{equation}
where $H_{\mu \nu}= \partial_{\mu}S_{\nu}- \partial_{\nu}S_{\mu}$
is the conformal gauge field strength.
With the help of  Eqs.(\ref{cri1}) and (\ref{cur1}), we obtain
\begin{equation}
\sqrt{g}R(\Omega)=\sqrt{g}R(\{\}) +
6\sqrt{g}(\nabla_{\alpha}S^{\alpha}-S_{\alpha}S^{\alpha}),
\end{equation}
where $\nabla_{\alpha}$ is the ordinary covariant derivative in Riemann space.

Under  the  conformal transformations, the  scalar  field
in 4-dimensions transforms as follow;
\begin{equation}
    \phi'(x) = exp(-\Lambda)\phi(x).
\end{equation}
Finally, the conformally invariant Lagrangian function $e\phi^{2}R(\Omega)$
up to total derivatives can be expressed as follows;
\begin{equation}
     \sqrt{g}\phi^{2}R(\Omega) = \sqrt{g}\phi^{2}R(\{\}) -
6\sqrt{g}\phi^{2}S_{\alpha}S^{\alpha}
     - 6\sqrt{g}S^{\alpha}\partial_{\alpha}\phi^{2}.
\label{div}
\end{equation}

Defining the conformally covariant derivative $D_{\alpha}$,
\begin{equation}
D_{\alpha}\phi= \partial_{\alpha}\phi + S_{\alpha}\phi,
\end{equation}
we have the following expression of  the conformally invariant
induced gravity action in terms of  $g_{\alpha \beta}, S_{\alpha}$ and $\phi$;
\begin{equation}
      S =\int d^{4}x \sqrt{g}
      [-\frac{\xi}{2}R(\Omega)\phi^{2} +
 \frac{1}{2}D_{\alpha}\phi D^{\alpha}\phi -
 \frac{1}{4}H_{\alpha\beta}H^{\alpha\beta} -
 \frac{\lambda}{4!}\phi^{4}],
\label{gen}
\end{equation}
where we have  excluded the curvature square terms.
The parameter $\xi$
and $\lambda$ are dimensionless constants.
Using Eq.(\ref{div}) we can rewrite this action  in
terms of Riemann curvature scalar $R(\{\})$;
\begin{equation}
      S=\int d^{4}x \sqrt{g}
      [-\frac{\xi}{2}R(\{\})\phi^{2} +
 \frac{1}{2}\partial_{\alpha}\phi\partial^{\alpha}\phi-
 \frac{1}{4}H_{\alpha \beta}H^{\alpha \beta}
+(1+6\xi) S^{\alpha}(\partial_{\alpha}\phi)\phi+
 \frac{1}{2}(1+6\xi)S_{\alpha}S^{\alpha}\phi^{2}-
 \frac{\lambda}{4!}\phi^{4}].
\label{gen1}
\end{equation}
Here we are interested in the $\xi$ range, $-\frac{1}{6} < \xi < 0$.

\section{The One-loop Effective Potential in De Sitter Space}

In this section, we have found the one-loop effective potential
of the above action in De Sitter space
using the background field method and zeta-function
regularization in Ref.\cite{Dewitt,Allen}.

We consider the quantum fluctuations of the scalar field
$\phi$ and the torsion field $S_{\alpha}$, and treat the
metric $g_{\mu\nu}$ as a classical background field;

\begin{equation}
g_{\mu \nu}=g_{\mu \nu}^{b},~~~ S_{\mu}=\tilde{S_{\mu}},~~~
\phi= \phi_{b} + \tilde{\phi}.
\end{equation}
Let us expand the action  Eq.(\ref{gen1}) around the background fields,
then we have the quadratic action of the quantum fluctuations;
\begin{equation}
     I_{2} =\int d^{4}x\sqrt{g}[
-\frac{1}{2} \xi R(\{\}) \tilde{\phi}^{2}
+ \frac{1}{2}\partial_{\alpha}\tilde{\phi} \partial^{\alpha}\tilde{\phi}
-\frac{\lambda}{4}\phi_{b}^{2}\tilde{\phi}^{2}
-\frac{1}{4}\tilde{H}_{\mu\nu} \tilde{H}^{\mu \nu}
-(1+6\xi)\phi_{b}\nabla_{\alpha}\tilde{S}^{\alpha}\tilde{\phi}
+ \frac{1}{2}(1+6\xi)\tilde{S}_{\alpha}\tilde{S}^{\alpha}\phi_{b}^{2}].
\label{iq}
\end{equation}
For the sake of convenience, we define a potential $\tilde{V}$ as follows;
\begin{equation}
    \tilde{V}  \equiv
	 \frac{1}{2} \xi R(\{\}) \phi^{2} +
    \frac{\lambda}{4}\phi^{4}
= \tilde{V}(\phi_{b}) + \frac{1}{2} \tilde{V}^{''}(\phi_{b})
    \tilde{\phi}^{2}.
\end{equation}

By the Hodge decomposition theorem, one form $ S^{\mu} $ in De Sitter space
can be decomposed into two parts,
a co-closed form $ S_{T}^{\mu} $ and an exact form $ S_{L}^{\mu} $
because there is no harmonic one-form;
\begin{equation}
S^{\mu}= S^{\mu}_{T}+ S^{\mu}_{L},
\end{equation}
where the co-closed form
$ S^{\mu}_{T} $ satisfies
$ \nabla_{\mu}S^{\mu}_{T}=0, $
and the exact form $ S^{\mu}_{L} $ can be written as
$S_{L}^{\mu}=\partial^{\mu}\chi$ for a function $\chi$.
The above decomposition is orthogonal;
\begin{equation}
\int d^{4}\sqrt{g} S^{\mu}_{T}S_{L}^{\nu}=0.
\end{equation}
Choosing the following gauge fixing term,
\begin{equation}
   \triangle I_{2} = \frac{1}{2}\alpha \int d^{4}x \sqrt{g}
(\nabla_{\mu}\tilde{S}_{L}^{\mu}+\alpha^{-1}(1+6\xi)\phi_{b}\tilde{\phi})^{2},
\end{equation}
and  adding the gauge fixing term to the quadratic action Eq.(\ref{iq})
whose independent quantum fields are
$ \{ \tilde{S}_{L}^{\mu}, \tilde{S}_{T}^{\mu}, \tilde{\phi} \}$,
we have the following gauge fixed quadratic action of the quantum fluctuations;
\[
     I_{2} + \triangle I_{2} = \frac{1}{2} \int d^{4}x \sqrt{g} [
\tilde{S}^{\mu}_{T}(-\Box_{\mu\nu}+R_{\mu\nu}+(1+6\xi)\phi^{2}_{b}
     g_{\mu\nu})\tilde{S}^{\nu}_{T}
\]
\[
+ \alpha \tilde{S}^{\mu}_{L}(-\nabla_{\mu} \nabla_{\nu}
 + \alpha^{-1}(1+6\xi)\phi_{b}^{2}g_{\mu\nu})\tilde{S}_{L}^{\nu}
\]
\begin{equation}
+ \tilde{\phi}(
-\Box - \tilde{V}''(\phi_{b}) +\alpha^{-1}(1+6\xi)^{2}\phi_{b}^{2}
) \tilde{\phi}].
\end{equation}
The one-loop generating functional in the landau gauge in which $\alpha$ goes
to
the infinity ($\alpha \rightarrow \infty$) is
\begin{equation}
Z_{1}=[\frac{det\{\mu^{-2}Q\}}{det\{\mu^{-2}(W+(1+6\xi)\phi_{b}^{2})\}~
det\{\mu^{-2}(Q-\tilde{V}''(\phi_{b}))\}}]^{1/2},
\end{equation}
where $ Q=-\Box $ without zero mode, $ W=-\Box+\frac{\bar{R}}{4}$,
and $ \bar{R} $ is the constant scalar curvature in De Sitter space.
We have dropped the spurious zero mode integrations in the path integral
because the zero mode of conformal
factor can be absorbed into the fixed constant background of dilaton field.

One-loop effective potential for the quantum fluctuations of the torsion vector
and the scalar field in Coleman-Weinberg
sector \cite{Coleman} (we assume that $\lambda$ is order of $(1+6\xi)^{2}$)
can be obtained using the
zeta-function regularization \cite{Allen,Gillkey};
\begin{equation}
\tilde{V}_{1}(\phi)=\tilde{V}(\phi)
+\frac{1}{2\Omega}\ln det\{\mu^{-2}(W+(1+6\xi)\phi_{b}^{2})\},
\end{equation}
where $ \Omega=\frac{8\pi^{2}a^{4}}{3} $ is the volume of
De Sitter space, and $\mu$ is a parameter with mass dimension.

In the large radius limit, $ (1+6\xi)a^{2}\phi^{2} >> 1 $, the above
effective potential becomes
\begin{equation}
\tilde{V}_{1}(\phi)=\tilde{V}(\phi)+\frac{3(1+6\xi)^{2}}{64\pi^{2}} \phi^{4}
(\ln \frac{(1+6\xi)\phi^{2}}{\mu^{2}}-\frac{3}{2})
+\frac{(1+6\xi)}{64\pi^{2}}\bar{R}\phi^{2}
(\ln \frac{(1+6\xi)\phi^{2}}{\mu^{2}}-1).
\label{effpot}
\end{equation}

\section{Semiclassical Equation of Motion}

In this section we will analyze the semiclassical equation of motion
for the scalar field and the metric considering the effective one-loop
potential which has been obtained in the previous section.
We have found the effective Lagrangian density as follows;
\begin{equation}
\sqrt{g}L_{eff}=
\sqrt{g}[-\frac{1}{2}\xi R(\{\})\phi^{2}
+\frac{1}{2}g^{\alpha\beta}\partial_{\alpha}\phi\partial_{\beta}\phi
-V_{eff}(\phi)],
\label{effl}
\end{equation}
where
\begin{equation}
V_{eff}(\phi,a) =
 \frac{\lambda}{4!}\phi^{4}+
\frac{ 3(1+6\xi)^{2}\phi^{4}}{64\pi^{2}}
(\ln \frac{(1+6\xi)\phi^{2}}{\mu^{2}}-\frac{3}{2})
+\frac{(1+6\xi)}{64\pi^{2}}\bar{R}\phi^{2}
(\ln \frac{(1+6\xi)\phi^{2}}{\mu^{2}}-1)+\rho_{v},
\end{equation}
Here we have shifted the vacuum energy by $\rho_{v}$ which might be attributed
to quantum corrections of other fields we have not considered.
By varying the action Eq.(\ref{effl}), we get  two  equations of motion
for the scalar field and the metric;
\begin{equation}
\Box \phi
+\xi R(\{\}) \phi
= - \frac{\partial V_{eff}}{\partial\phi},
\label{box}
\end{equation}
\begin{equation}
\xi \phi^{2}(R_{\mu\nu}-\frac{1}{2}g_{\mu\nu}R)=
\partial_{\mu}\phi\partial_{\nu}\phi
-\frac{1}{2} g_{\mu\nu} \partial_{\alpha}\phi\partial^{\alpha}\phi
-\xi(g_{\mu\nu}\Box\phi^{2}-\nabla_{\mu}\partial_{\nu}\phi^{2})
+g_{\mu\nu}V_{eff}(\phi),
\label{long}
\end{equation}
where we have not considered the backward contribution of the curvature
dependence of the effective potential into the Einstein equation (\ref{long}).

To  investigate  the  symmetry-breaking  equation  in this
model,  let us  look for  the  solution of these equations with $\phi=\sigma=$
constant.
The scalar equation of motion Eq.(\ref{box}) is reduced to the following;
\begin{equation}
\xi R(\{\}) =
-\frac{1}{\sigma} \frac{\partial V_{eff}}{\partial \phi} \mid_{\phi=\sigma}.
\label{c}
\end{equation}
The trace of Einstein Eq.(\ref{long}) is
\begin{equation}
-\xi R(\{\})\phi^{2}
+\partial_{\alpha}\phi\partial^{\alpha}\phi
-4 V_{eff}
+3\xi\Box\phi^{2}
=0,
\label{tra}
\end{equation}
which implies for the constant $\phi=\sigma$
\begin{equation}
\xi R(\{\}) = -\frac{4}{\sigma^{2}} V_{eff}(\sigma).
\label{tra1}
\end{equation}
With the help of Eqs.(\ref{c}) and (\ref{tra1}), we have the
symmetry-breaking equation;
\begin{equation}
(\frac{\partial V_{eff}}{\partial \phi}- \frac{4V_{eff}(\phi)}{ \phi})
\mid_{\phi=\sigma}=0.
\label{break}
\end{equation}
Therefore,  the symmetry breaking equation for the induced gravity
is different from the usual
$ \frac{\partial V_{eff}}{\partial \phi}
\mid_{\phi=\sigma}=0 $ in the scalar theory
with Einstein-Hilbert action.

Presently, it is assumed that we are in the broken symmetry phase  with
$\phi=\sigma$. If $V_{eff}(\sigma) \neq 0$,
this uniform background energy density acts like a
cosmological  constant in Einstein equation. By the requirement of
the vanishing of the cosmological constant in the true vacuum of flat
space, the constant part of $V_{eff}(\sigma,a)$ can be determined
\begin{equation}
\rho_{v}=\frac{3(1+6\xi)^{2}\sigma^{4}}{128\pi^{2}}.
\label{xyx}
\end{equation}
{}From the Eq.(\ref{break}), we can express the parameter $\mu$
in terms of $\sigma$ as follows;
\begin{equation}
\ln \frac{(1+6\xi)\sigma^{2}}{\mu^{2}} =
1-\frac{8\pi^{2}\lambda}{9(1+6\xi)^{2}}.
\label{xyz}
\end{equation}

In De Sitter space, the metric can be written as
\begin{equation}
     ds^{2} = dt^{2} - e^{2Ht}d\vec{x}^{2},
\end{equation}
and the scalar curvature is $\bar{R} = -12H^{2}$.
Using Eqs.(\ref{xyx}) and (\ref{xyz}), the effective potential
Eq.(\ref{effpot}) for the dilaton field in De Sitter becomes
\begin{equation}
\tilde{V}_{eff}(\phi_{s}) =
\frac{3}{64\pi^{2}} \phi_{s}^{4}(\ln \phi_{s}^{2} -\frac{1}{2})
+ \frac{3}{128\pi^{2}}
- \frac{3}{2} H^{2} \phi_{s}^{2}
  (\frac{1}{8\pi^{2}}\ln \phi_{s}^{2} - \frac{\lambda_{s}}{9})
-\frac{6\xi}{(1+6\xi)}H^{2}\phi_{s}^{2}~,
\label{fep}
\end{equation}
where, for the sake of convenience, we have defined
\begin{equation}
\phi_{s}\equiv\sqrt{(1+6\xi)}\phi, ~~~
\sigma_{s}\equiv\sqrt{(1+6\xi)}\sigma, ~~~
\lambda_{s}\equiv\frac{\lambda}{(1+6\xi)^{2}},
\end{equation}
and chosen the unit $\sigma_{s}=1$.
This effective potential governs the evolution of the dilaton field in De
Sitter
space through the equation of motion
\begin{equation}
\Box \phi = -\frac{\partial\tilde{V}_{eff}(\phi)}{\partial\phi}.
\end{equation}
It is found that the effective potential (\ref{fep}) shows a phase transition
which is sensible for an inflationary scenario.
The critical radius $1/H$ of the phase transition has been obtained at
$\frac{1}{H} \cong 19$ in the plotting of the
one-loop effective potential (\ref{fep})
varying the radius $\frac{1}{H}$ with
the fixed $\lambda_{s} = 1.0$ and $(1+6\xi) = 0.1$.

The combination of Eqs.(\ref{box}) and (\ref{tra}) gives
\begin{equation}
\phi\Box\phi+\partial_{\alpha}\phi\partial^{\alpha}\phi
+\phi\frac{\partial V_{eff}}{\partial\phi} -4 V_{eff}
+3\xi\Box\phi^{2}
=0.
\label{xx}
\end{equation}
{}From the assumption that $\phi$ is spatially  homogeneous,
the above Eq.(\ref{xx}) is reduced to
\begin{equation}
(1+6\xi)(\ddot{\phi} + 3H\dot{\phi}  +\frac{\dot{\phi}^{2}}{\phi})
+(V'_{eff}(\phi)-\frac{4}{\phi}V_{eff}(\phi)) =0.
\label{tmp5}
\end{equation}
When $\xi = -\frac{1}{6}$,  the  induced gravity  Lagrangian  is
consistent only if the form of the effective potential $V_{eff}(\phi)$ is
quartic.
Therefore, the spontaneous symmetry breaking is impossible for $\xi =
-\frac{1}{6}$ in De Sitter space.
The trace of Einstein Eq.(\ref{tra}) becomes
\begin{equation}
12\xi H^{2}\phi^{2}+\dot{\phi}^{2}
+6\xi(\phi \ddot{\phi} + \dot{\phi}^{2} + 3H\dot{\phi}\phi)
-4V_{eff}(\phi)
=0.
\label{tmp6}
\end{equation}
We are interested in the inflationary solutions of Eqs.(\ref{tmp5}) and
(\ref{tmp6}),  where the expansion rate H is very large in comparison with
other  quantities, and  scalar  field  changes slowly (slow
rollover) \cite{Accetta,Kaiser};
\begin{equation}
|\frac{\dot{\phi}}{\phi}| << H, ~~~|\ddot{\phi}| << 3H|\dot{\phi}|,
{}~~|{\dot{\phi}}^{2}| << V_{eff}(\phi).
\end{equation}
In the slow-rolling inflationary regime,  Eqs.(\ref{tmp5}) and (\ref{tmp6})
are reduced to the followings;
\begin{equation}
3H\dot{\phi}_{s} =
\frac{4~}{\phi_{s}} V_{eff}(\phi_{s}) - V^{'}_{eff}(\phi_{s})~,
\label{tmp8}
\end{equation}
\begin{equation}
H^{2} = \frac{(1+6\xi)}{3\xi \phi_{s}^{2}}V_{eff}(\phi_{s})~,
\label{tmp9}
\end{equation}
where
\begin{equation}
V_{eff}(\phi_{s}) =
\frac{3}{64\pi^{2}} \phi_{s}^{4}(\ln \phi_{s}^{2} -\frac{1}{2})
+ \frac{3}{128\pi^{2}}
- \frac{3}{2} H^{2} \phi_{s}^{2}
  (\frac{1}{8\pi^{2}}\ln \phi_{s}^{2} - \frac{\lambda_{s}}{9})~.
\end{equation}
In the slow-rolling phase, the contribution from the
$\frac{4V_{eff}(\phi_{s})}{\phi_{s}}$ part of the driving term
in the right-hand side of
Eq.(\ref{tmp8}) should be nearly equal to the contribution of the
$V^{'}_{eff}(\phi_{s})$ term so that the dilaton field could roll down slowly
compared with the expansion rate $H$.
This slow rolling inflationary phase surely can not
happen at the very center of the potential, but near the origin such that
\begin{equation}
\ln\phi_{s}^{2} \cong 8\pi^{2}(\frac{\lambda_{s}}{9}-\frac{2\xi}{(1+6\xi)}).
\end{equation}

When the scalar filed $\phi_{s}$ reaches $\phi_{s} \cong 1$, it is expected
that
the dilaton field oscillates
about the true vacuum with damping because
the  dilaton field  can be coupled to other matter fields through  Yukawa
couplings $Tr \bar{\psi} \Gamma(\phi \psi)$).
Through this dissipation process, the vacuum energy density of the  symmetric
phase, $\frac{3\sigma_{s}^{4}}{128\pi^{2}}$,
is eventually converted into radiation and matters.

\section{Conclusion}

     We have considered that the Newton's gravitational  constant
$G_{N}$  is generated through the spontaneous symmetry breaking
of  a  conformal symmetry.  It  is  possible to formulate the conformally
induced gravity in Riemann space.
However, the  spontaneous symmetry breaking via radiative  correction
does not work for a scalar field with $\xi=-\frac{1}{6}$.
     We  have extended minimally the Riemann  space  to
Riemann-Cartan  space  to incorporate  the
torsion vector which  is effectively playing the role  of  a  conformal
gauge  field,  then  the  dimensionless  coupling  constant $\xi$ is
arbitrary.
With the introduction of the conformal gauge field, the mechanism
of spontaneous symmetry breaking via radiative correction does
work as in the case of the massless scalar electrodynamics.
The  computation of the one-loop effective potential is performed
by  zeta-function  regularization in  De  Sitter  space.
Considering  this effective potential, we have analyzed the
semi-classical equation  of motion  of the dilaton field.
We will consider the case of non-vanishing torsion background and
the detail analysis of the effective potential within the context of
inflation scenario later.

\noindent
{\it Acknowledgments:} This work was supported in part by the Ministry of
Education through BSRI-93-206, the Center for Theoretical Physics of SNU
and Hanyang University.

\end{document}